\documentstyle[preprint,prb,aps,epsfig,amssymb,amsbsy,amsmath,multirow]{revtex}
\newcommand{\op}[1]{%
    \fontdimen12\textfont3=2pt\fontdimen12\scriptfont3=1.4pt%
    \!\null\mathop{\vphantom{#1}\smash{#1}}\limits_{\sim}\null\!}

\newcommand{\vek}[1]{{\!\vec{\,#1}}}

\newcommand{\figref}[1]{Fig.~\protect\ref{#1}}
\newcommand{\fmref}[1]{(\protect\ref{#1})}

\newcommand{\Tr}{\text{Tr}}
\newcommand{\EinsOp}
           {\;\smash{\raisebox{-1.1ex}{$\!\!\stackrel{\!\mbox{1}
            \hspace{-0.4ex}\rule[0.0ex]{0.06ex}{1.60ex}}{\sim}$}}}
\begin{document}
\tightenlines \draft 
\title{Heisenberg exchange parameters of molecular magnets from
the high-temperature susceptibility expansion}
\author{Heinz-J\"urgen Schmidt and J\"urgen Schnack\cite{correspondence}}
\address{Universit\"at Osnabr\"uck, Fachbereich Physik \\  
         Barbarastr. 7, 49069 Osnabr\"uck, Germany}
\author{Marshall Luban}
\address{Ames Laboratory \& Department of Physics and Astronomy,
Iowa State University\\ Ames, Iowa 50011, USA}
\maketitle
\begin{abstract}
We provide exact analytical expressions for the magnetic
susceptibility function in the high temperature expansion for
finite Heisenberg spin systems with an arbitrary coupling
matrix, arbitrary single-spin quantum number, and arbitrary
number of spins. The results can be used to determine unknown
exchange parameters from zero-field magnetic susceptibility
measurements without diagonalizing the system Hamiltonian.  We
demonstrate the possibility of reconstructing the exchange
parameters from simulated data for two specific model
systems. We examine the accuracy and stability of the proposed
method.

\end{abstract}
\widetext
\section{Introduction and summary}

Due to advances of synthesis methods in coordination chemistry
\cite{Cor:NATO96,MSG:AC01} and, especially, in polyoxometalate
chemistry,\cite{MPP:CR98,MSS:ACIE99,Chi:PRL00,Mue:CPC01} the
field of molecular magnets has undergone rapid progress in
recent years. To date each of the identical molecular magnets
contains as few as two and up to several dozens of paramagnetic
ions (``spins").  The largest paramagnetic molecule synthesized,
the polyoxometalate \cite{MSS:ACIE99} \{Mo$_{72}$Fe$_{30}$\}
contains 30 iron ions of spin $s=5/2$.  Although these materials
are synthesized as macroscopic samples, i.e. crystals or
powders, the intermolecular magnetic interactions are generally
utterly negligible as compared to the intramolecular
interactions.  Therefore, measurements of their magnetic
properties reflect mainly ensemble properties of single
molecules. It appears that in most cases the magnetic properties
of these molecules are rather well described by the Heisenberg
model with possibly small anisotropy
corrections.\cite{BeG90,WSK:IO99,DGP:IC93,CCF:IC95,PDK:JMMM97}

In order to calculate from first principles the energy spectrum
and the magnetic properties it is essential to accurately
determine the exchange parameters, as these measure the strength
of the two-spin interactions. The usual procedure consists of
the following steps: diagonalization of the system Hamiltonian
for an arbitrary choice of exchange parameters; calculation of
the partition function and a quantity, such as the
magnetization, which can be compared to experimental data;
variation of the exchange parameters so as to achieve an optimal
fit between the experimental and theoretical data.  This can be
a rather straightforward task for simple configurations such as
small spin rings with a single nearest-neighbor coupling.
However, for more complicated systems, for example where two or
more exchange parameters may be needed it may be impractical to
perform such calculations for different trial choices of the
parameters, even if one is guided by chemical and structural
information. Worse yet, for large spin systems, where in
practice the Hamiltonian cannot be diagonalized, it may even be
impossible to fix the exchange parameters, since one cannot
calculate the weak-field susceptibility over the full
tempearature range, for comparison with experiment.

In this article we explore the possibility of determining the
exchange parameters of Heisenberg molecular magnets using the
exactly calculable first three terms of the high-temperature
expansion of the zero-field susceptibilty $\chi(\beta)$, in
powers of $\beta$. In comparison to earlier work
\cite{TaS:JPSJ97,BEU:EPJB00} exact results are derived here for
these three coefficients, for arbitrary coupling, arbitrary
single-spin quantum number, and an arbitrary number of
spins. The procedure we explore here opens up the possibility of
treating large systems where the Hamiltonian can no longer be
diagonalized. We also outline the necessary algorithms for
calculating additional terms of the high-temperature expansion,
but these would require computer algebra methods.

Deducing exchange parameters from experimental susceptibility
data of course presumes that one possesses accurate estimates of
the number of paramagnetic ions, the spectroscopic splitting
factor(s), diamagnetic contributions, temperature-independent
paramagnetism, etc. We do not consider these issues here;
instead we concentrate on the question, with what accuracy
exchange parameters can be deduced using theoretical
susceptibility data, for a Heisenberg molecular magnet, limited
to the high-temperature regime. We do not want to mix questions
about the accuracy of the method with uncertainties in
experimental data which are of different origin.

This article is organized as follows. After recalling basic
definitions in Sec.~\ref{sec-2} we derive the trace formulae
which are necessary for the high-temperature expansion in
Sec.~\ref{sec-3}. These results are summarized in
Sec.~\ref{sec-4} and related to the expansion coefficients of
the Taylor series of $\chi(\beta)$.  The accuracy and stability
of the proposed method are illustrated for two selected model
systems in Sec.~\ref{sec-5}. We briefly summarize our findings
in Sec.~\ref{sec-6}.

\section{Definitions}
\label{sec-2}

For the following investigations we assume that the spin system
under consideration is well described by the following general
bilinear Heisenberg Hamilton operator
\begin{eqnarray}\label{1}
\op{H}_0
\equiv
\sum_{\mu < \nu}^{}\,J_{\mu\nu} 
\op{\vek{s}}_\mu \cdot \op{\vek{s}}_\nu
=
\sum_{\mu<\nu} J_{\mu\nu} 
\sum_{i=1}^{3} \op{s}_\mu^{(i)}\op{s}_\nu^{(i)}
\ .
\end{eqnarray}
The exchange parameters $J_{\mu\nu}$ are given in units of energy, with
$J_{\mu\nu}<0$ resulting in ferromagnetic, $J_{\mu\nu}>0$ in
antiferromagnetic coupling.  $\op{\vek{s}}_\mu$ are the
dimensionless single-particle spin operators and
$\op{s}_\mu^{(i)}$ denotes the spin components
$i\in\{1,2,3\}\equiv\{x,y,z\}$ of the spin at site $\mu$. The
interaction with the magnetic field is taken into account by
including the Zeeman term
\begin{eqnarray}\label{3}
\op{H}
\equiv 
\op{H}_0
+
g \mu_B B \op{S}^{(3)}
\ ,
\end{eqnarray}
where $\op{S}^{(3)}$ is the $z$-component of the total spin
operator, and $g$ is the spectroscopic splitting factor. For
simplicity we are supposing that all of the paramagnetic ions
possess the same value of $g$. The present results can be easily
generalized to cases where several different paramagnetic
species are present. As usual the magnetisation is defined as
\begin{eqnarray}\label{5}
{\mathcal M}
&=&
-
\frac{1}{Z}\,
\Tr\left\{g \mu_B \op{S}^{(3)}\, \mbox{e}^{-\beta\op{H}}\right\}
\ ,\ 
Z
=
\Tr\left\{\exp(-\beta \op{H})\right\}
\ ,
\end{eqnarray}
$Z$ being the partition function. Accordingly the zero-field
magnetic susceptibility is
\begin{eqnarray}\label{6}
\chi_0
&=&
\left(\frac{\partial {\mathcal M}}{\partial B}\right)_{B=0}
=
\beta (g \mu_B)^2
\frac{\Tr\left\{\op{S}^{(3) 2}\, \exp(-\beta \op{H}_0)\right\}}
     {\Tr\left\{\exp(-\beta \op{H}_0)\right\}}
\ .
\end{eqnarray}
The susceptibility may be renormalised by division of $(g
\mu_B)^2$
\begin{eqnarray}\label{7}
\hat{\chi}(\beta)\equiv\frac{\chi_0(\beta)}{(g \mu_B)^2}
\ .
\end{eqnarray}
A Taylor expansion of $\hat{\chi}$ in powers of $\beta$,
starting from (\ref{6}), yields after some calculations
\begin{eqnarray}  
\label{17}
\hat{\chi}(\beta) 
= 
\sum_{n=1}^\infty c_n \beta^n
&=& 
\frac{\mu_0}{t_0}\beta -\frac{\mu_1}{t_0}\beta^2 +
\left(-\frac{t_2 \mu_0}{2 t_0^2}+\frac{\mu_2}{2
t_0}\right)\beta^3
\\
& & 
+\left(\frac{t_3 \mu_0}{6 t_0^2}+\frac{t_2 \mu_1}{2 t_0^2}-
\frac{\mu_3}{6 t_0}\right)\beta^4
\nonumber
\\
& & 
+\left(\mu_0\left(\frac{t_2^2}{4 t_0^3}-\frac{t_4}{24 t_0^2}\right)
-\frac{t_3 \mu_1}{6 t_0^2}-\frac{t_2 \mu_2}{4
t_0^2}+\frac{\mu_4}{24 t_0}\right)\beta^5
\nonumber
\\
& & +\ldots
\nonumber
\ ,
\end{eqnarray}
where
\begin{eqnarray}\label{8}
t_n\equiv \Tr\left\{ \op{H}_0^n\right\}
\ ,\quad 
\mu_n
\equiv 
\Tr\left\{\op{H}_0^n \op{S}^{(3) 2} \right\}
\ ,\quad
n\in{\Bbb N},
\ .
\end{eqnarray}
Comparing expansion \fmref{17} with a Curie-Weiss law
\begin{eqnarray}
\label{18}
\hat{\chi}_{CW}(\beta)   
&=& 
\frac{\hat{C}}{T-T_c}
= 
\hat{C}k_B\beta (1-k T_c \beta)^{-1}
=
\hat{C}k_B\beta \left(1+k_B T_c \beta
+(k_B T_c \beta)^2
+
\ldots\right)
\end{eqnarray}
yields the known theoretical results \cite{CWL} for the Curie
constant $\hat{C}$ and the Weiss temperature $T_c$
\begin{eqnarray}\label{19}
k_B \hat{C}
=
\frac{\mu_0}{t_0}
\quad \text{and}\quad
k_B T_c = -\frac{\mu_1}{\mu_0}
\ .
\end{eqnarray}
Generally, the $\beta^3$-term of \fmref{18} will not coincide
with the $\beta^3$-term of the rigorous expansion \fmref{17}.
This is due to the fact that the Curie-Weiss law itself is only
valid at high temperatures.\cite{AsM} Thus, the $\beta^3$-term
of \fmref{17} constitutes a first (quantum) correction to the
Curie-Weiss law for lower temperatures.

\section{Trace formulae}
\label{sec-3}

Expanding the terms occuring in the trace \fmref{8}
\begin{eqnarray}\label{10}
t_n
=
\Tr\left\{ \op{H}_0^n\right\}
&=&
\Tr\left\{ \sum_{\mu<\nu} J_{\mu\nu} 
\sum_{i=1}^{3} \op{s}_\mu^{(i)}\op{s}_\nu^{(i)}\right\}^n
\end{eqnarray}
and using $\Tr (A\otimes B)=(\Tr A)(\Tr B)$ one ends up with
terms of the form 
\begin{eqnarray}\label{11}
\Tr  (A_1 \ldots A_\ell), 
A_\nu\in\{\op{s}^{(1)},\op{s}^{(2)},\op{s}^{(3)} \}
\ .
\end{eqnarray}
Here the spin operators $\op{s}$ without a site index denote
operators operating in the single-spin Hilbert space ${\Bbb
C}^{2s+1}$.
Let $\ell_i (i=1,2,3)$ denote the number of occurrences of
$\op{s}^{(i)}$ in the product $A_1\ldots A_\ell$.  One can
easily show that $\Tr (A_1\ldots A_\ell)$ is non-zero only if
all $\ell_i$ are even or all $\ell_i$ are odd. We give a list of
the simplest cases, where the trace is non-zero:
\begin{eqnarray}\label{12}
\ell=0 & : & \Tr\left\{\EinsOp \right\} = 2s+1,\\
\ell=2 & : & \Tr\left\{\op{s}^{(i)2} \right\}  =\sum_{m=-s}^s m^2 = \frac{1}{3}s(s+1)(2s+1),\nonumber\\
\ell=3 & : & \Tr\left\{\op{s}^{(1)}\op{s}^{(2)}\op{s}^{(3)} \right\}  
         = - \Tr\left\{\op{s}^{(3)}\op{s}^{(2)}\op{s}^{(1)} \right\}
         = \frac{i}{6}s(s+1)(2s+1),\nonumber\\
\ell=4 & : & \Tr\left\{\op{s}^{(i)4} \right\} =  \sum_{m=-s}^s m^4 = \frac{1}{15}s(s+1)(2s+1)(3s^2+3s-1),\nonumber\\
\ell=4 & : & \Tr\left\{\op{s}^{(1)2}\op{s}^{(2)2} \right\}   =  \frac{1}{30}s(s+1)(2s+1)(2s^2+2s-1)
\nonumber
\ .
\end{eqnarray}
Note that there are no $\ell=1$ cases; hence all traces 
$\Tr\left\{\op{s}_{\mu_1}^{(i_1)j_1} 
\ldots \op{s}_{\mu_K}^{(i_K)j_K}\right\}$ vanish 
if at least for some $k=1,\ldots,K$ the site index
$\mu_k$ is different from the other $\mu_k'$ and $j_k=1$,
therefore,
\begin{eqnarray}
t_1 
=
\Tr\left\{ \op{H}_0\right\}
=
0
\ .
\end{eqnarray}
From the list \fmref{12} we obtain for $n=2$:
\begin{eqnarray}\label{13}
t_2
=
\Tr\left\{ \op{H}_0^2\right\}
&=&
\sum_{\mu<\nu} J_{\mu\nu}^2 
\sum_{i=1}^3
\Tr\left\{\op{s}_\mu^{(i)2}\otimes
\op{s}_\nu^{(i)2}\otimes{{\EinsOp}}_{N-2}\right\}
\\
&=&
\left( \sum_{\mu<\nu} J_{\mu\nu}^2\right) \frac{1}{3} s^2
(s+1)^2 (2s+1)^N
\nonumber
\ ,
\end{eqnarray}
where ${{\EinsOp}}_{N-2}$ denotes the unit operator in the
Hilbert space of $N-2$ spins.  For $n=3$ there occur two kinds
of non-zero terms:
\begin{eqnarray}\label{14}
\Tr\left\{
\op{s}_\mu^{(1)} \op{s}_\mu^{(2)}\op{s}_\mu^{(3)}
\otimes \op{s}_\nu^{(1)} \op{s}_\nu^{(2)}\op{s}_\nu^{(3)}
\otimes{{\EinsOp}}_{N-2} \right\}
\end{eqnarray}
and those terms obtained by permutations of $\{ 1,2,3\}$, and
\begin{eqnarray}\label{15}
\Tr\left\{
\op{s}_\mu^{(i)2}\otimes \op{s}_\nu^{(i)2}
\otimes \op{s}_\kappa^{(i)2} \otimes{{\EinsOp}}_{N-3} \right)
\ ,
\end{eqnarray}
where $\mu,\nu,\kappa$ are pairwise distinct and $i=1,2,3$.
Consequently we obtain
\begin{eqnarray}\label{16}
t_3
=
\Tr\left\{ \op{H}_0^3\right\}
&=&
T_1 + T_2
\ , \\
T_1 
&=&
\sum_{\mu<\nu} J_{\mu\nu}^3\,
3!(2s+1)^{N-2} 
\Tr\left\{\op{s}^{(1)} \op{s}^{(2)}\op{s}^{(3)}\right\}^2
\\
&=&
- \sum_{\mu<\nu} J_{\mu\nu}^3 \frac{1}{6} s^2 (s+1)^2 (2s+1)^N
\nonumber\ , \\
T_2 
&=&
3!\left( \sum_{\mu<\nu<\kappa} J_{\mu\nu}
J_{\mu\kappa}J_{\nu\kappa} \right) 
(2s+1)^{N-3}
\sum_{i=1}^3
\Tr\left\{\op{s}^{(i)2} \right\}^3
\\
&=&
\left( \sum_{\mu<\nu<\kappa} J_{\mu\nu}
J_{\mu\kappa}J_{\nu\kappa} \right)
\frac{2}{3}s^3(s+1)^3(2s+1)^N
\nonumber
\ .
\end{eqnarray}

\section{The Taylor expansion of $\chi$}
\label{sec-4}

\subsection{The linear term}

For the linear term we just need $t_0$ and $\mu_0$. $t_0$ is
nothing else than the dimension of the $N$-spin Hilbert space 
\begin{eqnarray}\label{22}
t_0
=
\Tr\left\{ \EinsOp\right\}
=
(2s+1)^N
\ .
\end{eqnarray}
For $\mu_0$ we obtain with the help of list \fmref{12}
\begin{eqnarray}\label{23}
\mu_0 
&=& 
\Tr\left\{ \op{S}^{(3)2}\right\}
= 
\frac{1}{3} \Tr\left\{ \op{\vek{S}}^2 \right\}
\\
&=& 
\frac{1}{3} \left(\sum_\nu 
\Tr\left\{ \op{\vek{s}}_\nu^2\right\} 
+
\sum_{\nu\neq\mu} 
\Tr\left\{ \op{\vek{s}}_\nu\cdot\op{\vek{s}}_\mu\right\}
\right)
=
\frac{1}{3} N s(s+1) t_0
\nonumber
\ ,
\end{eqnarray}
and therefore
\begin{eqnarray}\label{23-2}
c_1
&=& 
\frac{1}{3} N s(s+1)
\ .
\end{eqnarray}
Hence, combining \fmref{19}, \fmref{22}, and \fmref{23} we
find that the Curie-Weiss constant is given by
\begin{eqnarray}\label{24}
k_B C 
&\equiv& 
k_B (g \mu_B)^2 \hat{C}
=
(g \mu_B)^2   \frac{1}{3} N s(s+1)
\ ,
\end{eqnarray}
which is identical to the mean-field result.\cite{CWL,AsM}

\subsection{The $\beta^2$-term}

For the quadratic term we calculate $\mu_1$ directly
\begin{eqnarray}\label{25}
\mu_1 
&=&
\Tr\left\{\op{H}_0 \op{S}^{(3) 2} \right\}
=
\frac{1}{3}
\Tr\left\{ \op{H}_0\op{\vek{S}}^2 \right\}
\\
&=&
\frac{1}{3}
\sum_{\mu < \nu, i} J_{\mu \nu} 
\Tr\left\{
\op{s}_\mu^{(i)} \op{s}_\nu^{(i)}  
\left(
2 \sum_{\kappa < \lambda,j}
(\op{s}_\kappa^{(j)} \op{s}_\lambda^{(j)})
+ N s(s+1) \EinsOp
\right)
\right\}
\nonumber
\ .
\end{eqnarray}
Since $\Tr\{\op{s}_\mu^{(i)} \op{s}_\nu^{(j)}\}=0$ for
$\mu\neq\nu$ the second summand vanishes. According to section
\ref{sec-3}, the first one is only non-zero if $\mu=\kappa,
\nu=\lambda, i=j$. Hence
\begin{eqnarray}\label{26}
\mu_1 
&=&
\frac{2}{3} \sum_{\mu < \nu, i }  J_{\mu \nu} 
\Tr\left\{
\op{s}_\mu^{(i)2} \op{s}_\nu^{(i)2}
\right\}
= 2 \left(  \sum_{\mu < \nu}  J_{\mu \nu}  \right)
\left(\frac{1}{3} s (s+1)\right)^2 (2s+1)^N
\\
&=&
\frac{2}{9} 
\left(  \sum_{\mu < \nu}  J_{\mu \nu}  \right)
\, s^2 (s+1)^2 t_0
\nonumber
\ .
\end{eqnarray}
This results in a coefficient
\begin{eqnarray}\label{26-2}
c_2=  -\frac{2}{9} 
\left(  \sum_{\mu < \nu}  J_{\mu \nu}  \right) 
\, s^2 (s+1)^2
\end{eqnarray}
and a Weiss temperature of
\begin{eqnarray}\label{27}
k_B T_c 
&=& 
- \frac{\mu_1}{\mu_0}
=
- \frac{2}{3 N} s (s+1)
\left(  \sum_{\mu < \nu}  J_{\mu \nu}  \right)
\ ,
\end{eqnarray}
which agrees with the mean field result.\cite{CWL,AsM}

\subsection{The $\beta^3$-term and higher orders}

Using the same method the next term can be calculated. The
coefficient reads
\begin{eqnarray}\label{28}
c_3 
&=&
\frac{2}{27} s^3 (s+1)^3
\sum_{\kappa < \lambda < \mu}\left(J_{\kappa\lambda} J_{\kappa\mu}
+J_{\kappa\mu}J_{\lambda\mu}+J_{\kappa\lambda}J_{\lambda\mu}\right)
\\
\nonumber
&& 
-\frac{1}{18} s^2 (s+1)^2 \sum_{\kappa < \lambda}
\left(J_{\kappa\lambda}^2 \right)
\nonumber
\ .
\end{eqnarray}
Hence $c_3$ depends not on all exchange parameters but only on a
certain combination of the second moments (or correlation
coefficients) of the exchange parameters.

It is clear how coefficients of higher orders are to be
calculated, although the expressions rapidly become very
lengthy. Computer algebra methods would have to be employed to
determine and evaluate those expressions.

\section{Applications}
\label{sec-5}

The first two expansion coefficients in the Taylor expansion
\fmref{17} of the susceptibility in powers of $\beta$ that
involve the exchange parameters are $c_2$ and $c_3$.  We propose
exploiting accurate numerical estimates for $c_2$ and $c_3$ to
provide information on the exchange parameters. The following
version of \fmref{17}
\begin{eqnarray}  \label{29}
k_B\,T \left(k_B\, T \hat{\chi} - c_1 \right)
= 
\sum_{n=2}^\infty c_n \beta^{n-2}
=
c_2
+
c_3 \beta
+ \cdots
\ ,
\end{eqnarray}
is especially convenient since we can determine $c_2$ and $c_3$
by a least-squares fit between a low degree polynomial in
$\beta$ to measured values of the quantity $k_B\,T \left(k_B\, T
\hat{\chi} - c_1 \right)$, compare also Ref.~\onlinecite{TaS:JPSJ97}. To
increase the accuracy of the determined values of $c_2$ and
$c_3$ it is generally advisable to use a polynomial in $\beta$
of degree two or higher. In an iterative procedure one could
check the extent to which the values of $c_2$ and $c_3$ change
as higher degree polynomials are used.  We remind the reader
that we assume that the spectroscopic splitting factor $g$,
which is necessary to obtain function \fmref{29}, has accurately
been determined, e.g. by ESR measurements.

In the following two examples we test the proposed method for
two model systems. Starting with given exchange parameters we
determine the partition function by exact diagonalization of the
Hamiltonian and calculate the susceptibility. Our aim is to
extract accurate estimates for the exchange parameters
from a limited number of susceptibility data points. We also
assess how the method depends on the chosen temperature
interval and how robust the method is when the experimental data
includes random error. 

If the system has two exchange parameters $J_1$ and $J_2$ it is
obvious from \fmref{26-2} and \fmref{28} that $c_2$ is a linear
function of $J_1$ and $J_2$ whereas $c_3$ is a quadratic
function of these variables. Hence $J_1$ and $J_2$ can be
determined by numerically solving a simple non-linear equation
once accurate values of $c_2$ and $c_3$ are available. In cases
where multiple solutions exist additional magneto-chemical
information, for instance concerning the sign of the exchange
parameters, must be considered.

A similar complication arises for the special class of so-called
isospectral spin systems, where the same eigenvalue spectrum is
associated with a host of different choices of the exchange
parameters.\cite{ScL:JPA01} For such systems it is in principle
impossible to obtain unique exchange parameters from
susceptibility data.

\subsection{Cubic spin array}

\begin{figure}[ht!]
\begin{center}
\epsfig{file=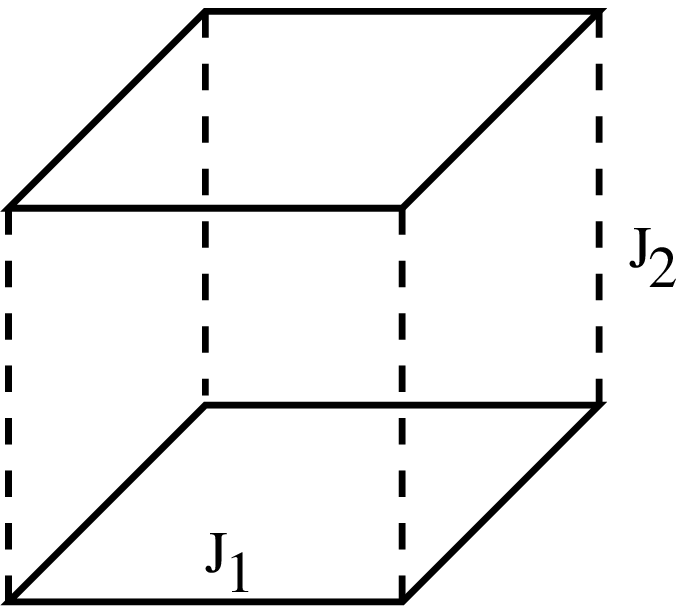,width=40mm}
\vspace*{1mm}
\caption[]{Structure of a spin cube of $N=8$ and $s=1$.  The
assumed exchange parameters are $J_1/k_B=20$~K, $J_2/k_B=10$~K.}
\label{F-1}
\end{center} 
\end{figure}

\begin{figure}[ht!]
\begin{center}
\epsfig{file=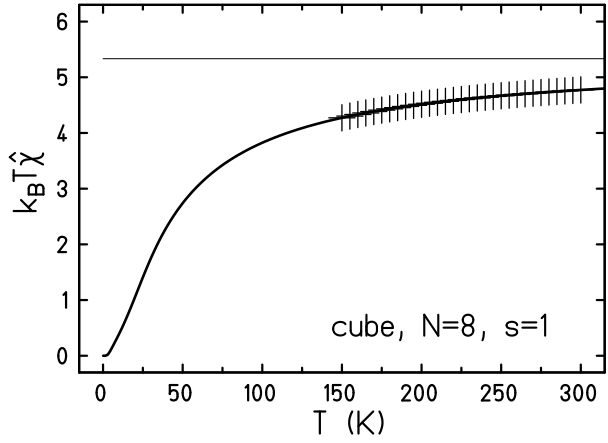,width=80mm}
\vspace*{1mm}
\caption[]{$k_B T \hat{\chi}$ data  (thick line) for the
spin cube shown in \figref{F-1}.  The data set used in the
calculations is denoted by crosses. The thin line is the
high temperature limit.}
\label{F-2}
\end{center} 
\end{figure}

\begin{figure}[ht!]
\begin{center}
\epsfig{file=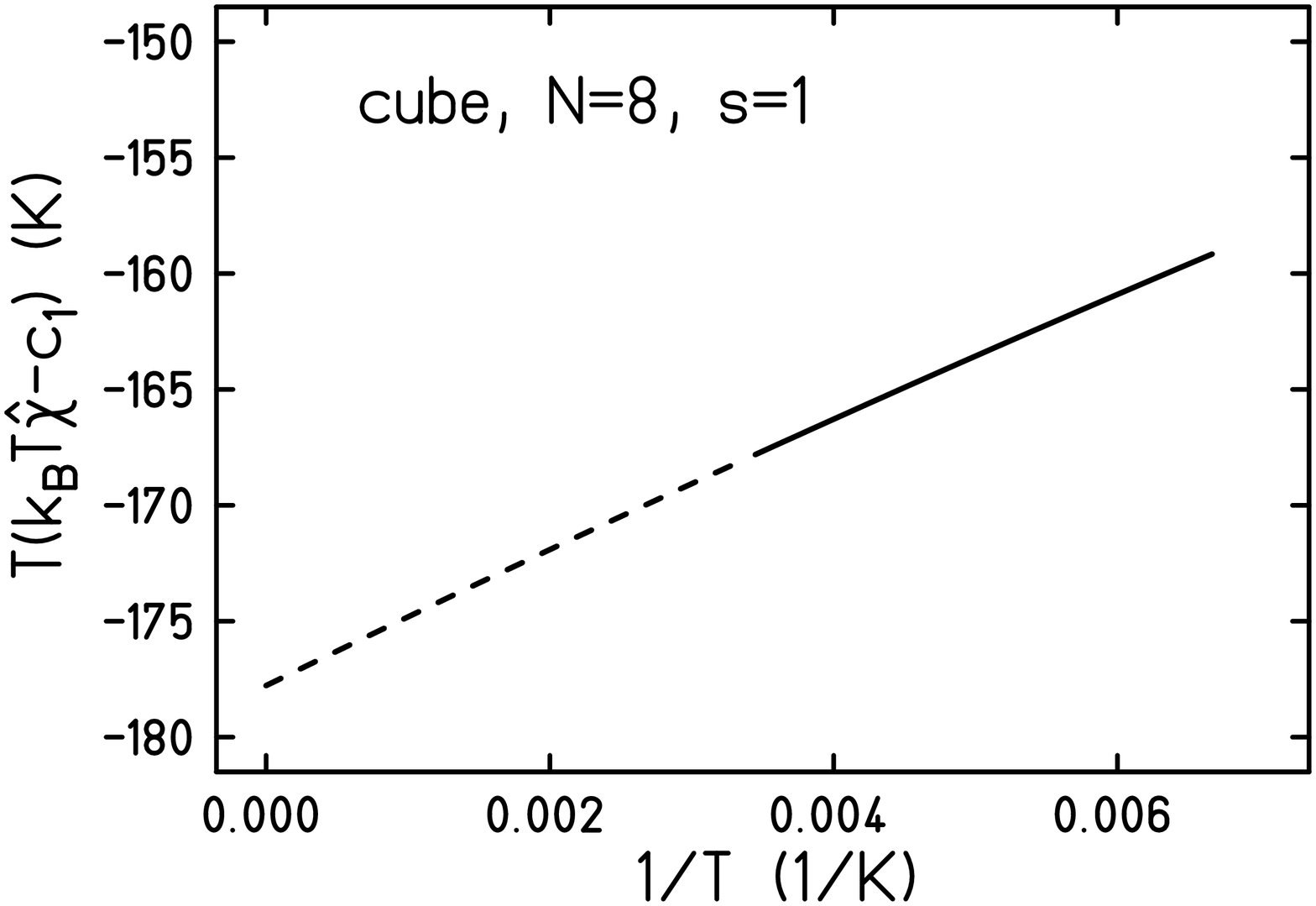,width=80mm}
\vspace*{1mm}
\caption[]{Functional dependence of 
$T \left( k_B T \hat{\chi} - c_1 \right)$
on $1/T$ (thick line)
for the data set (crosses) shown in \figref{F-2}.
The continuation  of this function for higher temperatures is
given by the  dashed line.}
\label{F-3}
\end{center} 
\end{figure}

Our first model system is a spin array where $N=8$ spins $s=1$
are located at the vertices of a cube. The spins are assumed to
interact via two antiferromagnetic exchange parameters of
typical magnitude, $J_1/k_B=20$~K between nearest neighbors in
the top and bottom face and $J_2/k_B=10$~K connecting the two
faces, see \figref{F-1}. The calculated magnetic susceptibility
is shown in \figref{F-2} as a thick line. The thin horizontal
line corresponds to the high-temperature limiting value of $k_B
T \hat{\chi}$ given by the coefficient $c_1$, see \fmref{17} and
\fmref{23-2}. For the determination of the underlying exchange
parameters a restricted set of susceptibility data is used. This
set consists of data taken in steps of 5~K from 150~K to 300~K
(crosses in \figref{F-2}). As already mentioned, not the
susceptibility $\hat{\chi}$ itself or $T\hat{\chi}$ are fitted
but 
\begin{eqnarray}  \label{30}
f(T)
&=&
T \left( k_B T \hat{\chi} - c_1 \right).
\end{eqnarray}
The graph of the latter function is shown in \figref{F-3}. One
notes that its dependence on $1/T$ in the relevant temperature
interval is already rather linear. Therefore one expects good
results using the fitting procedure based on a low degree
polynomial. Indeed, approximating $T \left(k_B T \hat{\chi} -
c_1 \right)$ by a polynomial of third order yields
$J_1/k_B=19.96$~K and $J_2/k_B=10.09$~K, a result which is not
improved by using polynomials of higher degree.

We also investigated how the chosen temperature interval
influences the analysis. Including only data between 150~K and
200~K yields exchange parameters which deviate by at most 10~\%
from the correct values.

\begin{figure}[ht!]
\begin{center}
\epsfig{file=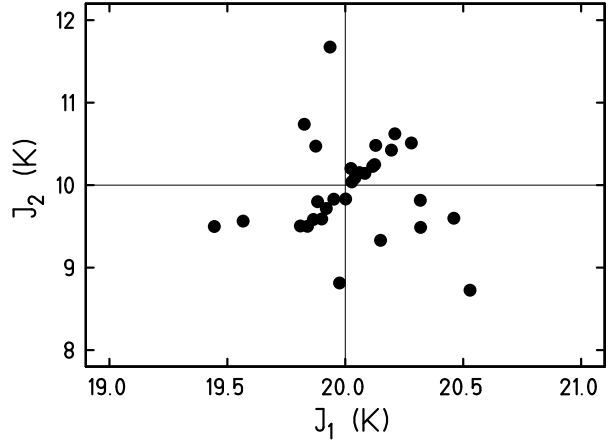,width=80mm}
\vspace*{1mm}
\caption[]{Distribution of exchange parameters of the spin cube
under the assumption that the susceptibility data incorporates
random errors at the level of $5$~\%.}
\label{F-4}
\end{center} 
\end{figure}

Different from our model investigation, experimental data
necessarily incorporates random noise. In order to estimate this
effect, the determination of the exchange parameters was carried
out with a number of modified data sets.  For each data set all
susceptibility values were allowed to deviate by at most $\pm
5$~\% from the correct values. The resulting exchange parameters
are shown in \figref{F-4}, each point corresponding to a different
modified data set. It appears that the influence of the
fluctuations is different for the two exchange parameters due to
the non-linear dependence of $J_1$ and $J_2$ on the expansion
coefficients $c_n$.

\subsection{Ring spin array}

\begin{figure}[ht!]
\begin{center}
\epsfig{file=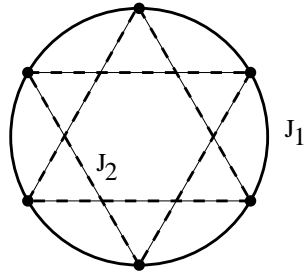,width=40mm}
\vspace*{1mm}
\caption[]{Structure of a
spin ring of $N=6$ and $s=5/2$ with nearest-neighbor $J_1$ and
next-nearest-neighbor coupling $J_2$. The assumed exchange parameters
are (a) $J_1/k_B=20$~K, $J_2=0$ and (b) $J_1/k_B=20$~K,
$J_2/k_B=10$~K.}
\label{F-5}
\end{center} 
\end{figure}

\begin{figure}[ht!]
\begin{center}
\epsfig{file=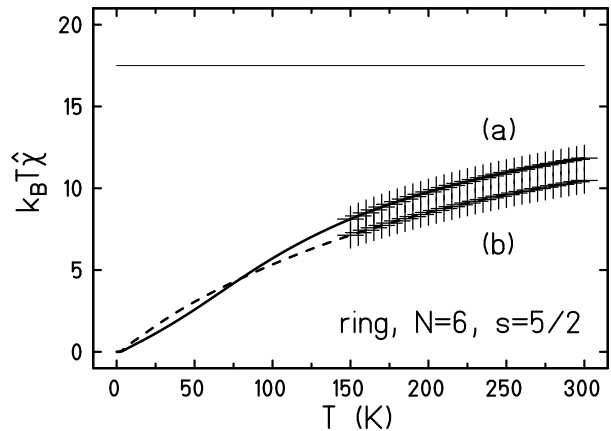,width=80mm}
\vspace*{1mm}
\caption[]{$k_B T \hat{\chi}$ data for the spin ring
introduced in \figref{F-5}. The thick curve shows case (a) and
the dashed curve shows case (b).  The thin line corresponds to
the high temperature limit.  The data sets used in the
calculations are depicted by crosses.}
\label{F-6}
\end{center} 
\end{figure}

\begin{figure}[ht!]
\begin{center}
\epsfig{file=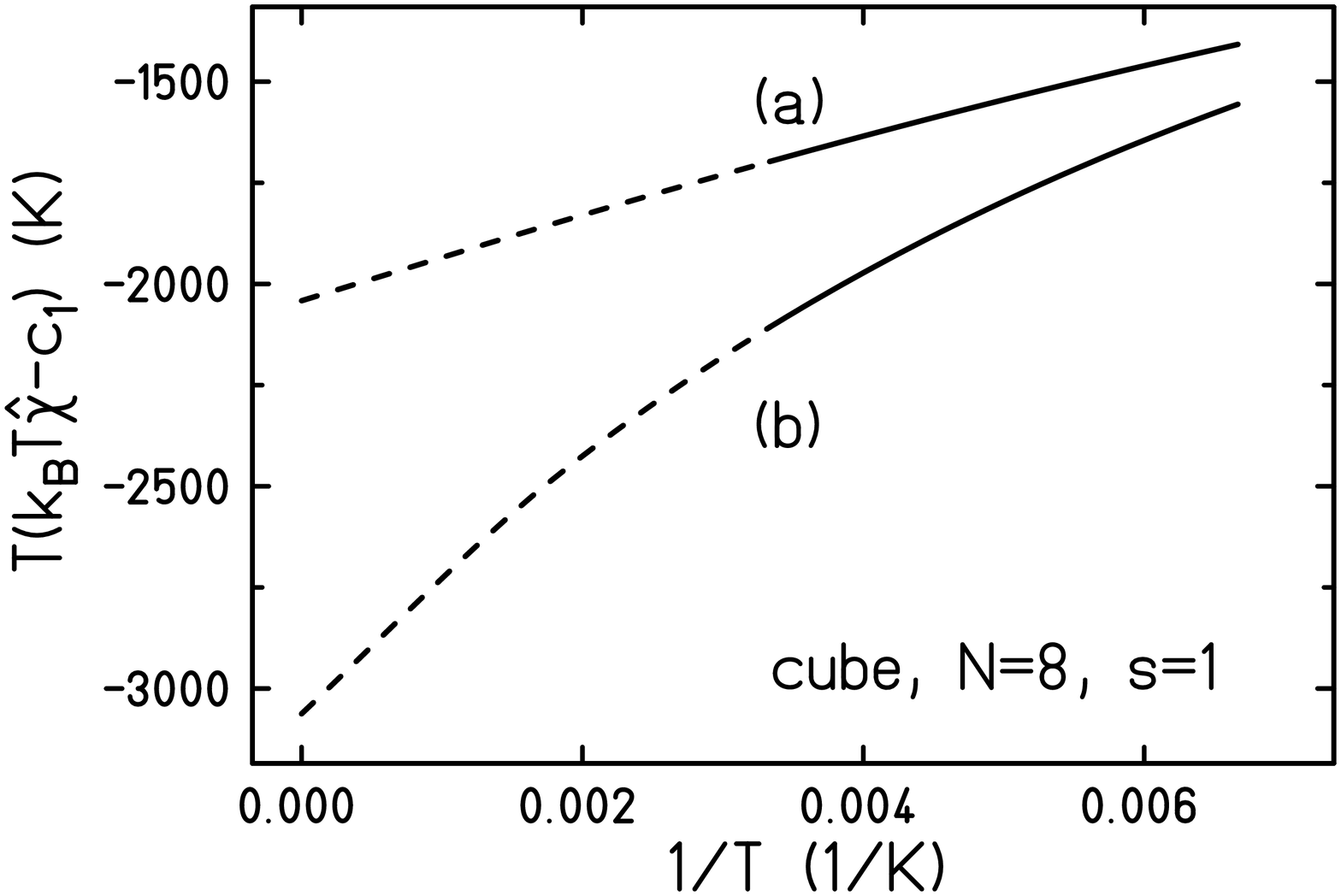,width=80mm}
\vspace*{1mm}
\caption[]{Functional dependence of 
$T \left( k_B T \hat{\chi} - c_1 \right)$
on $1/T$ (thick line)
for the data set (crosses) shown in \figref{F-6}.
The continuation  of these functions for higher temperatures is
given by dashed lines.}
\label{F-7}
\end{center} 
\end{figure}

As a second model system we investigate a ring with $N=6$ and
$s=5/2$. In order to further quantify the accuracy of the
proposed method, we chose two cases, case (a) where the spins
are connected only by a nearest-neighbor interaction
$J_1/k_B=20$~K, and case (b) where we have an additional
next-nearest-neighbor interaction $J_2/k_B=10$~K, see
\figref{F-5}. The corresponding values of $T\chi$, again
obtained by complete diagonalization of the Hamiltonian, are
given in \figref{F-6}. The thick curve corresponds to case (a)
and the dashed curve corresponds to case (b).  The thin line
again gives the high temperature limit. Data points used for our
analysis are depicted by crosses, again within the range between
150 K and 300 K. Note that for this model system the data points
deviate from the high temperature behavior much more than in the
previous example, i.e. the dependence of $f(T)$ on $1/T$
(\figref{F-7}) in the above temperature range especially for case
(b), is not as linear as it was in the previous example.

The present method is sufficiently robust to provide accurate
estimates of the exchange parameters without exploiting any
prior information. This is illustrated by using two exchange
parameters for case (a) and finding that $J_1/k_B=20.04$~K and
$J_2/k_B=-0.04$~K, which is in very good agreement with the
actual exchange parameters in case (a). For case (b) we find
$J_1/k_B=19.33$~K and $J_2/k_B=10.69$~K, which is not as
accurate as in the previous examples.  Nevertheless, even in
this worst case, using these exchange parameters results in $k_B
T \hat{\chi}$ which, when plotted in \figref{F-6}, is
indistinguishable for all temperatures from that for the correct
values of $J_1$ and $J_2$. Improved exchange parameters might be
achievable if measured data for $\chi$ is available for higher
temperatures, but unfortunately many substances -- especially
those containing solvent molecules for instance -- become
structurally unstable at such temperatures.

\section{Summary}
\label{sec-6}

In summary, we have provided exact analytical expressions for
the first three terms of the high-temperature expansion of the
magnetic susceptibility function for Heisenberg spin systems
with an arbitrary coupling matrix $J_{\mu\nu}$ and arbitrary
intrinsic spins $s$. The results can be rather easily used in
order to determine unknown exchange parameters for molecular
magnets from zero-field magnetic susceptibility data with good
accuracy, thereby obviating the need for diagonalizing the
system Hamiltonian.


\section*{Acknowledgments}

H.-J.~S.~would like to thank members of the Condensed Matter
Physics Group of Ames Laboratory for their warm hospitality
during a visit when a part of this work was performed. We thank
P.~K\"ogerler for helpful discussions. We also acknowledge the
financial support of a travel grant awarded jointly by NSF-DAAD.
Ames Laboratory is operated for the United States Department of
Energy by Iowa State University under Contract
No. W-7405-Eng-82.



\end{document}